\documentclass[twocolumn]{aastex631} 

\usepackage{graphicx}
\usepackage{amsmath} 
\usepackage{mathrsfs}
\usepackage{enumerate}
\usepackage{xspace}
\usepackage[T1]{fontenc}

\newcommand{\pycs}{{\tt PyCS3}\xspace}
\newcommand{\rxjeleven}{RXJ~1131$-$1231\xspace}

\newcommand{\qtwentytwo}{Q~2237$+$0305\xspace}

\newcommand{\invday}{days\ensuremath{^{-1}}\xspace}

\shorttitle{Quasar Microlensing Event Predictor}
\shortauthors{Fagin et al.}

\begin{document} 

\title{Predicting High-magnification Events in Microlensed Quasars in the Era of LSST using Recurrent Neural Networks}

\correspondingauthor{Joshua Fagin}
\email{jfagin@gradcenter.cuny.edu}

\author[0000-0001-8723-6136]{Joshua~Fagin}
\affiliation{The Graduate Center of the City University of New York, 365 Fifth Avenue, New York, NY 10016, USA}
\affiliation{Department of Astrophysics, American Museum of Natural History, Central Park West and 79th Street, NY 10024-5192, USA}
\affiliation{Department of Physics and Astronomy, Lehman College of the CUNY, Bronx, NY 10468, USA}

\author[0000-0002-4306-7366]{Eric~Paic}
\affiliation{Institute of Physics, Laboratory of Astrophysics, Ecole Polytechnique F\'ed\'erale de Lausanne (EPFL), Observatoire de Sauverny, 1290 Versoix, Switzerland}

\author[0009-0006-4977-6098]{Favio~Neira}
\affiliation{Institute of Physics, Laboratory of Astrophysics, Ecole Polytechnique F\'ed\'erale de Lausanne (EPFL), Observatoire de Sauverny, 1290 Versoix, Switzerland}

\author[0009-0009-6932-6379]{Henry~Best}
\affiliation{The Graduate Center of the City University of New York, 365 Fifth Avenue, New York, NY 10016, USA}
\affiliation{Department of Astrophysics, American Museum of Natural History, Central Park West and 79th Street, NY 10024-5192, USA}
\affiliation{Department of Physics and Astronomy, Lehman College of the CUNY, Bronx, NY 10468, USA}
\affiliation{Department of Theoretical Physics and Astrophysics, Faculty of Science,
Masaryk University, Kotlářská 2, CZ-611 37 Brno, Czech Republic}

\author[0000-0003-0930-5815]{Timo~Anguita}
\affiliation{Instituto de Astrofísica, Departamento de Ciencias Fisicas, Facultad de Ciencias Exactas, Universidad Andres Bello, Av. Fernandez Concha 700, Las Condes, Santiago, Chile}
\affiliation{Millennium Institute of Astrophysics, Nuncio Monseñor Sótero Sanz 100, Providencia, Santiago, Chile}

\author[0000-0001-7051-497X]{Martin~Millon}
\affiliation{Kavli Institute for Particle Astrophysics and Cosmology and Department of Physics, Stanford University, Stanford, CA 94305, USA}

\author[0009-0000-4476-5003]{Matthew~O'Dowd}
\affiliation{The Graduate Center of the City University of New York, 365 Fifth Avenue, New York, NY 10016, USA}
\affiliation{Department of Astrophysics, American Museum of Natural History, Central Park West and 79th Street, NY 10024-5192, USA}
\affiliation{Department of Physics and Astronomy, Lehman College of the CUNY, Bronx, NY 10468, USA}

\author[0000-0001-6116-2095]{Dominique~Sluse}
\affiliation{STAR Institute, Universit\'e de Li\`ege, Quartier Agora Allée du six Août, 19c B-4000 Liège, Belgium}

\author[0000-0001-8554-7248]{Georgios~Vernardos
}
\affiliation{The Graduate Center of the City University of New York, 365 Fifth Avenue, New York, NY 10016, USA}
\affiliation{Department of Astrophysics, American Museum of Natural History, Central Park West and 79th Street, NY 10024-5192, USA}
\affiliation{Department of Physics and Astronomy, Lehman College of the CUNY, Bronx, NY 10468, USA}

\begin{abstract}
Upcoming widefield surveys, such as the Rubin Observatory's Legacy Survey of Space and Time (LSST), will monitor thousands of strongly lensed quasars over a 10 yr period. Many of these monitored quasars will undergo high-magnification events (HMEs) through microlensing, as the accretion disk crosses a caustic --- places of infinite magnification. Microlensing allows us to map the inner regions of the accretion disk as it crosses a caustic, even at large cosmological distances. The observational cadences of LSST are not ideal for probing the inner regions of the accretion disk, so there is a need to predict HMEs as early as possible, to trigger high-cadence multiband or spectroscopic follow-up observations. Here, we simulate a diverse and realistic sample of 10 yr quasar microlensing light curves to train a recurrent neural network to predict HMEs before they occur, by classifying the locations of the peaks at each time step. This is the first deep-learning approach for predicting HMEs. We give estimates of how well we expect to predict HME peaks during LSST and benchmark how our metrics change with different cadence strategies. With LSST-like observations, we can predict approximately 55\% of HME peaks, corresponding to tens to hundreds per year and a false-positive rate of around 20\% compared to the total number of HMEs. Our network can be continuously applied throughout the LSST survey, providing crucial alerts for optimizing follow-up resources.
\end{abstract}

\keywords{Quasars(1319) --- Active galactic nuclei(16) --- Quasar microlensing(1318) --- Gravitational lensing(670) --- Neural networks(1933) --- Time series analysis(1916)}

\section{Introduction} \label{sec:introduction}

The most widely accepted unified model of active galactic nuclei (AGNs) postulates that their structure revolves around a central supermassive black hole (SMBH) at the centers of galaxies~\citep{antonucci93,urry95}. The tidal stretching and viscous friction experienced by matter in the immediate vicinity of the SMBH create hot plasma, forming an accretion disk that emits a bright continuum of light across a wide range of wavelengths with stochastic variability. Quasars are bright AGNs with unobscured accretion disks. Their radial energy profile is believed to follow the thin-disk model~\citep{ShakuraSunyaev, PageThorne74}, but recent observations from microlensing~\citep{Poindexter08, Poindexter10, Blackburne15,  Munoz16, Morgan18} and reverberation mapping~\citep{Mudd_2018, Jha_2022} have found larger accretion disk sizes by a factor of \mbox{$\sim$2--4}. Farther out, clouds of ionized gas form the broad-line regions (BLRs) and narrow-line regions, reverberating accretion disk light and inducing broad- and narrow-line emissions in the quasar spectra, depending on the gas velocity~\citep{Blandford82,Sluse07,Odowd15,Grier17}. Quasars also influence the evolution of their host galaxy, due to the energy outflow they create~\citep[e.g.,][]{Di_Matteo_2008,Fabian12,Kormendy_2013}. Quasars are among the most luminous objects in the Universe, making them valuable cosmological probes for measuring the expansion rate of the Universe~\citep{Lusso19, Wong20} and understanding the reionization mechanism~\citep{Grissom14}. These, however, require a precise understanding of quasar structure and the underlying physical processes.
 
Strongly lensed quasars occur when light from a quasar source is bent by a galaxy along the line of sight, forming multiple images of the quasar. Each image of the strongly lensed quasar can be independently magnified through microlensing due to stars or other compact objects in the lensing galaxy. This effect leads to a micromagnification of the light profile of the accretion disk. It is possible to measure the size of a microlensed accretion disk using single-epoch, multi-wave-band observations, using the different apparent sizes of the disk in different wave bands~\citep[e.g.,][]{Mediavilla15}. Alternatively, long-term monitoring of a strongly lensed quasar allows us to detect microlensing by measuring the micromagnification variation as the alignment between the microlenses and the background quasars varies over time. This long-term monitoring requires suppressing the intrinsic variability of the quasar by taking the pairwise difference of two strongly lensed images, corrected for the time delay between the images. If the intrinsic variability is perfectly subtracted, the resulting difference light curve, which we will refer to as the \emph{microlensing light curve}, only contains the microlensing signal. These microlensing light curves provide constraints not only on the size of the disk~\citep{Kochanek04, Morgan08, Cornachione2020a, Rivera2024} but also on the geometry of the BLR~\citep{Sluse11, sluse14, paic22, savic24} and can reveal the existence of substructures in the accretion disk, such as secondary SMBHs~\citep{Yan2014, Millon2022}. 

A high-magnification event (HME) occurs when a quasar approaches or crosses a microcaustic --- a region of infinite magnification. As a result, a specific area of the quasar's accretion disk becomes highly magnified. As the quasar moves relative to the microcaustic, the differential magnification provides a way of scanning through these different emission regions. In order to constrain the structure of the AGN engine through microlensing, however, we require high-cadence and multiwavelength observations. These microlensing events can last from as little as a few weeks to a few years. 
It has been shown that X-ray monitoring of the HME can place constraints on the spin, inclination, and size of the innermost stable circular orbit of the black hole~\citep[e.g.,][]{Chartas17}.
Furthermore, optical monitoring provides the opportunity to measure the temperature profile of the inner regions of the disk~\citep[e.g.,][]{Eigenbrod08}.
High-cadence and multiwavelength observations near caustic-crossing events are crucial for extracting the detailed accretion disk structure encoded in microlensing light curves~\citep[e.g., ][]{krawczynski2019,vernardos19, best2022resolving}. 
In the best-case scenario, the entire caustic-crossing event will be observed by expensive and high-cadence follow-up to place the most stringent constraints on the inner workings of the AGN. 
Therefore, being able to predict these events ahead of time is a critical step in order to trigger appropriate follow-up observations.

The most furnished microlensing light-curve data set to date was released by COSMOGRAIL~\citep{millon20} with 20 yr long microlensing light curves. Out of the 23 observed lensed quasars (17 doubly lensed and six quadruply lensed) and a total of 203 seasons of monitoring, only three microlensing light curves display microlensing events higher than 1~mag. These light curves only contain observations from the $r$ band, severely limiting the constraints that can be placed on the AGN structure and the ability to predict HMEs. 

The advent of widefield monitoring surveys, such as the Rubin Observatory's Legacy Survey of Space and Time (LSST), is expected to increase the sample of monitored lensed quasars at least tenfold. LSST is expected to simultaneously monitor hundreds to thousands of strongly lensed quasars~\citep{OguriMarshall2010} in six UV/optical wave bands (\textit{ugrizy}) at around 55--185 samplings per band or around 800 observations across the 10 yr~\citep{abell2009lsst,Prsa_2023}. Around 30\%--40\% of these monitored lensed quasars are expected to have strong enough intrinsic variability to precisely measure time delays between lensed images within the 10 yr for cosmological inference~\citep{Taak_2023}. Tens to hundreds of quasar images should undergo HMEs of more than 1~mag per year~\citep{Neira20}. We cannot accurately constrain the inner structure of the accretion disk using microlensing light curves at LSST cadences alone, due to the sparse and irregular sampling. However, LSST microlensing light curves will still allow us to detect the characteristic steady rise of a microlensing event occurring up to years before the peak. Furthermore, by using different wave bands, we can potentially determine the precise timing of the HME peak as it affects various regions of the disk, causing the wave bands to cross. 

An HME alert system was first developed for the Optical Gravitational Lensing Experiment~\citep{Wozniak_2000}, and a triggering mechanism for multiband follow-up observations was proposed by~\citet{Wyithe_2000}. These efforts focused on predicting HMEs in \qtwentytwo, due to its brightness and rapid microlensing time scale, but have not been used extensively. Thus far, there has been no prediction and follow-up of an HME as it unfolds. This is likely due to the small sample of lensed quasars and a lack of long-term, multiband monitoring that could distinguish microlensing from the intrinsic variability of the quasar. With LSST and other widefield surveys, however, this will no longer be the case. New and more sophisticated methods must be developed to be quickly and autonomously applied to the thousands of monitored lensed quasars expected in the near future. 

In this work, we aim to create a neural network able to predict the occurrence of microlensing HME peaks using the multiband microlensing light curves of LSST. This tool will be crucial for optimizing follow-up observations of such HMEs and caustic crossings. Machine learning was first applied to microlensed quasar light curves in~\citet{vernardos19} to measure the accretion disk size and temperature profile using a convolutional neural network for simulated data. \citet{best2022resolving}~also used a convolutional neural network to measure the black hole mass, inclination angle, and impact angle of simulated caustic-crossing events. As a pilot application, they measured the mass and inclination angle from archival caustic crossings of \qtwentytwo and found them to be consistent with previous estimates. Machine learning has also been extensively used to model intrinsic quasar variability~\citep[e.g., with simulated LSST quasar light curves, see][]{Fagin_2024}. Here, we train a recurrent neural network (RNN) using binary classification to identify the peaks of HMEs. Our RNN is able to account for the irregular cadences and seasonal gaps expected from LSST and could be continually applied to LSST microlensing light curves in real time as new data are obtained.

In Section~\ref{sec:sim}, we describe how we build our simulated data set of microlensing light curves. Section~\ref{sec:ML} presents our machine learning model architecture, training, and an example application to an LSST microlensing light curve. In Section~\ref{sec:results}, we benchmark the performance of our RNNs using different observational strategies, including the baseline case of simulated LSST-like observations. In Section~\ref{sec:discussion}, we discuss the results of our network and its applicability to LSST. Section~\ref{sec:conclusion} gives our concluding remarks. In this work, we assume a flat $\Lambda$CDM cosmology with $H_0 = 72.0$ $\text{km}\, \text{s}^{-1}\, \text{Mpc}^{-1}$, $\Omega_{m} = 0.26$, and $\Omega_{\Lambda} = 0.74$.

\section{Microlensing Light-curve Simulation} \label{sec:sim}

In order to train a machine learning model to predict HME peaks, we require a library of microlensing light curves that is representative of the expected LSST sample.
Since the goal is to predict the peaks of the HMEs identified on the light curves, we focus on constructing a library where the microlensing durations and amplitudes cover the range of what could be expected in current and future systems.
Each light curve in the training set has two main components: the microlensing signal and the associated correlated noise. The noise in the microlensing light curves is typically highly temporally correlated, often referred to as \emph{red noise} as opposed to uncorrelated Gaussian white noise~\citep{Millon2020, paic22}. The red noise is mainly due to imperfect subtraction of the intrinsic variability of the quasar or actual physical processes, such as the intrinsic variability being echoed by the BLR and differentially magnified by microlensing~\citep{sluse14,paic22}. Since these physical processes are not well understood and difficult to model, we rely on an empirical approach, where we treat all these effects as red noise added to our simulated microlensing light curves.

\subsection{Microlensing Simulation}
\label{subsec:microlensing_generation}
Microlensing brightness variations are uncorrelated between the multiple images of a strongly lensed quasar, unlike the intrinsic variability, which is identical in each image.
Since the latter can be subtracted, albeit imperfectly, leaving the so-called red noise (see the next section), we focus on simulating the microlensing variations.
To model the microlensing variability, we use the tool presented in \citet{Neira20}, which is able to generate simulated microlensing light curves for each image of the lensed quasars.

The tool takes as inputs an accretion disk, velocity, and macro lens model and outputs a library of light curves.
The light curves are generated as follows:
\begin{itemize}
    \item The accretion disk model is convolved with magnification maps selected from the GERLUMPH project \citep{Vernardos14}.
    \item In the convolved map, multiple tracks are defined to simulate the net transverse movement of the disk, accounting for the movement of the lens, microlenses, and observer, based on the velocity parameters and probing time.
    \item The pixel values along these tracks correspond to the microlensing light curves (see section~2 in~\citet{Neira20} for a description of the tool, models, and parameters).
\end{itemize}
We use the simulated microlensing light curves from Neira et al. (2024, in preparation), which makes use of the catalog of lensed quasars presented in~\cite{OguriMarshall2010}.
A total of 2821 mock systems are simulated.
While the details of the models and parameters used to generate the microlensing light curves from these systems can be found in Neira et al. (2024, in preparation), we briefly describe them here.

\begin{figure*}
    \centering
    \includegraphics[width=0.97\textwidth]{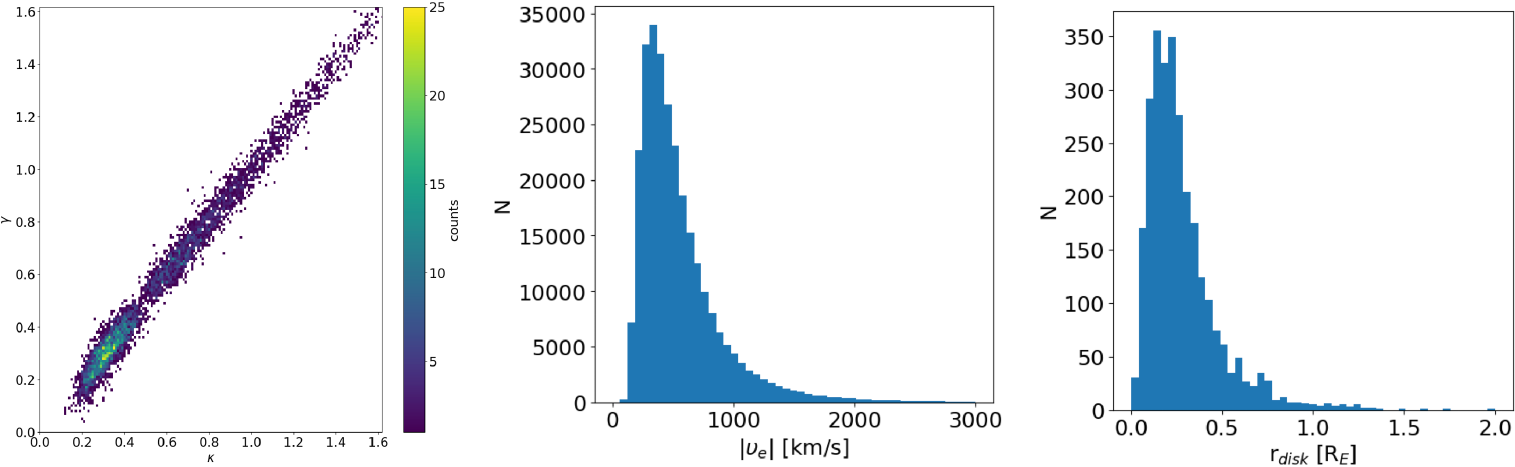}
    \caption{Distribution of the lensing macromodel parameters $\gamma$ and $\kappa$ (left), transverse velocity $v_e$ (middle), and accretion disk half-light} radius $r_{\text{disk}}$ (right).
    \label{fig:sim_param_dist}
\end{figure*}

The mass distribution of each lensing galaxy is modeled as a singular isothermal ellipsoid (SIE). For an SIE, the convergence and shear at the position of the quasar images are defined as
\begin{equation}
    \kappa(x, y) = \gamma(x,y) = \frac{1}{2} \frac{\theta_\text{E}\sqrt{q}}{\omega(x,y)} \, ,
\end{equation}
where $\theta_\text{E}$ is the Einstein radius, $q$ is the axis ratio of the lensing galaxy, and
$\omega(x,y)$ is the elliptical radius, where $x,y$ defines the coordinates of the lens plane, with the $x$-axis aligned with the major axis of the lens.
An additional term is added to the shear that corresponds to the lens environment, resulting in a small scatter in the final $\kappa$ and $\gamma$ distribution, shown in the left panel of Figure~\ref{fig:sim_param_dist}. The smooth matter fraction, $s$, is computed by assuming a light distribution and mass-to-light ratio of the lens \citep[see][and Neira et al. 2024, in preparation, for details]{Foxley-Marrable+2018,Vernardos2019a}.
This allows us to select a magnification map from the GERLUMPH library \citep{Vernardos14} corresponding to the selected $\kappa$, $\gamma$, and $s$.
Each map has 10,000 pixels per side, with a resolution of $0.025R_\text{E}$/pixel, where $R_\text{E}$ is the Einstein radius of the microlensing objects and is defined as
\begin{equation} \label{eq:einstein_radius}
    R_\text{E} = \sqrt{\frac{D_\text{S}D_\text{LS}}{D_\text{L}} \frac{4G  M }{c^2}} \, ,
\end{equation}
where $D_\text{L}$, $D_\text{S}$, and $D_\text{LS}$ are the angular diameter distances from the observer to lens, observer to source, and lens to source, respectively; $G$ is the gravitational constant; $c$ is the speed of light; and \mbox{$M=0.3 M_\odot$} is the mass of the microlenses (the approximate mean of a Salpeter initial mass function \citep{Salpeter1955}). 

The effective transverse velocity that defines the track along the magnification map is defined as
\begin{equation} \label{eq:velocity}
    \upsilon_\text{e} = \frac{\upsilon_\text{o}}{1+z_\text{l}} \frac{D_\text{LS}}{D_\text{L}} + \frac{\upsilon_\star}{1+z_\text{l}} \frac{D_\text{S}}{D_\text{L}} + \upsilon_\text{g} \, ,
\end{equation}
where $\upsilon_\text{o}$ is the transverse velocity of the observer, as measured with respect to the cosmic microwave background (CMB) velocity dipole \citep{Kogut1993}; $\upsilon_\star$ is the bulk velocity of the microlenses; and $\upsilon_\text{g}$ is the combined peculiar velocities of the lens and source.
The distribution of $\upsilon_\text{e}$ across all the light curves is shown in the middle panel of Figure~\ref{fig:sim_param_dist}.

The size of each accretion disk is assumed to follow a standard thin-disk model \citep{ShakuraSunyaev}:
\begin{equation}
    \label{eq:size}
    R_\lambda = 9.7 \times 10^{15}  \left(\frac{\lambda_\text{rest}}{\mu m}\right)^{4/3} \left(\frac{M_\text{BH}}{10^9M_\odot}\right)^{2/3} \left(\frac{f_E}{\eta}\right)^{1/3} [\text{cm}] \, ,
\end{equation}
where $R_\lambda$ is the size of the disk at $\lambda_\text{rest}$, $M_\text{BH}$ is the mass of the black hole, $f_\text{E}$ is the Eddington ratio, and $\eta$ is the accretion efficiency.
We adopted fixed typical values of $f_E=0.25$ and $\eta=0.15$ \citep[e.g.,][]{Blackburne11}.
For the black hole mass, we have adopted an empirical model from \cite{macleod10}:
\begin{equation}
    \log_{10}(M_\text{BH}) = 2 - 0.27 \cdot M_\text{i} \, ,
\end{equation}
where $M_\text{i}$ is the absolute magnitude of the source in the $i$ band.
Given a black hole mass estimate, the size of the accretion disk can be computed at any wavelength through Equation~(\ref{eq:size}). The influence of the specific shape of the brightness profile has only minor influence in the microlensing signal~\citep{Mortonson05,vernardos19}.
Instead, the signal is mostly influenced by the source size.
Thus, we adopt a Gaussian-shaped two-dimensional brightness profile, where we match the size from Equation~(\ref{eq:size}) to the half-light radii of the profile.
In the right panel of Figure~\ref{fig:sim_param_dist}, we show the distribution of this disk size, as measured at $6222$ $\text{\AA}$ in the rest frame, across all the systems with respect to their microlensing Einstein radius (see Equation~(\ref{eq:einstein_radius})).

With the above, we are able to generate microlensing light curves by subtracting the light curves between each pair of quasar images. For each pair of images for all simulated systems, we generate 100 light curves in the six LSST bands. The choice of which image to subtract (i.e., A$-$B versus B$-$A) is arbitrary, because the training label remains unchanged. The network does not inherently differentiate between a microlensing event in one image versus the other when analyzing the difference light curves. The two images can be directly subtracted, since we assume the time delay between each image has been accounted for, with any residual uncertainty incorporated as added red noise.

We produce a data set comprising 282,100 light curves. We randomly select 80\% of the light curves for training and reserve 10\% for validation and 10\% for testing.

\subsection{Red-noise Generation}
\label{subsec:rednoise}

\begin{figure*}
    \centering
    \includegraphics[clip, trim=0.2cm 0.2cm 0.5cm 0.2cm,width=\linewidth]{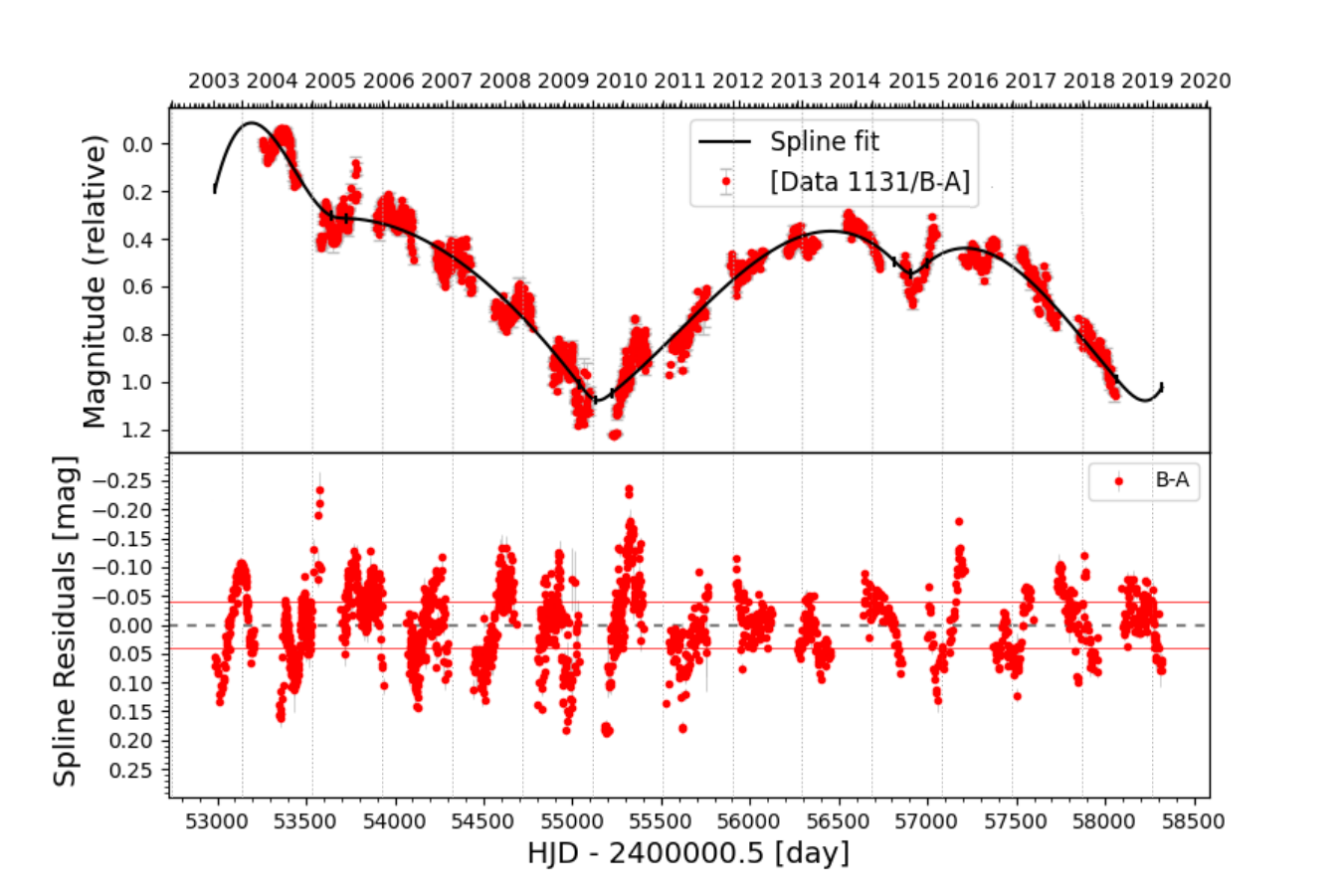}
    \caption{The top panel shows a COSMOGRAIL $r$-band microlensing light curve of \rxjeleven over a period of 18 yr, computed by subtracting image B from image A after shifting it by the time delay of 2.8 days~\citep{millon20}. A spline fit with $\eta=300$ days is applied to the microlensing light curve, and the black markers indicate the positions of the knots. The bottom panel displays the residuals of the spline fit, where the correlated noise is clearly visible. The horizontal dashed line and solid red lines indicate the median and standard deviation of the residuals, respectively.}
    \label{fig:lc_data}
\end{figure*}

\begin{figure*}
    \centering
    \includegraphics[width=\linewidth]{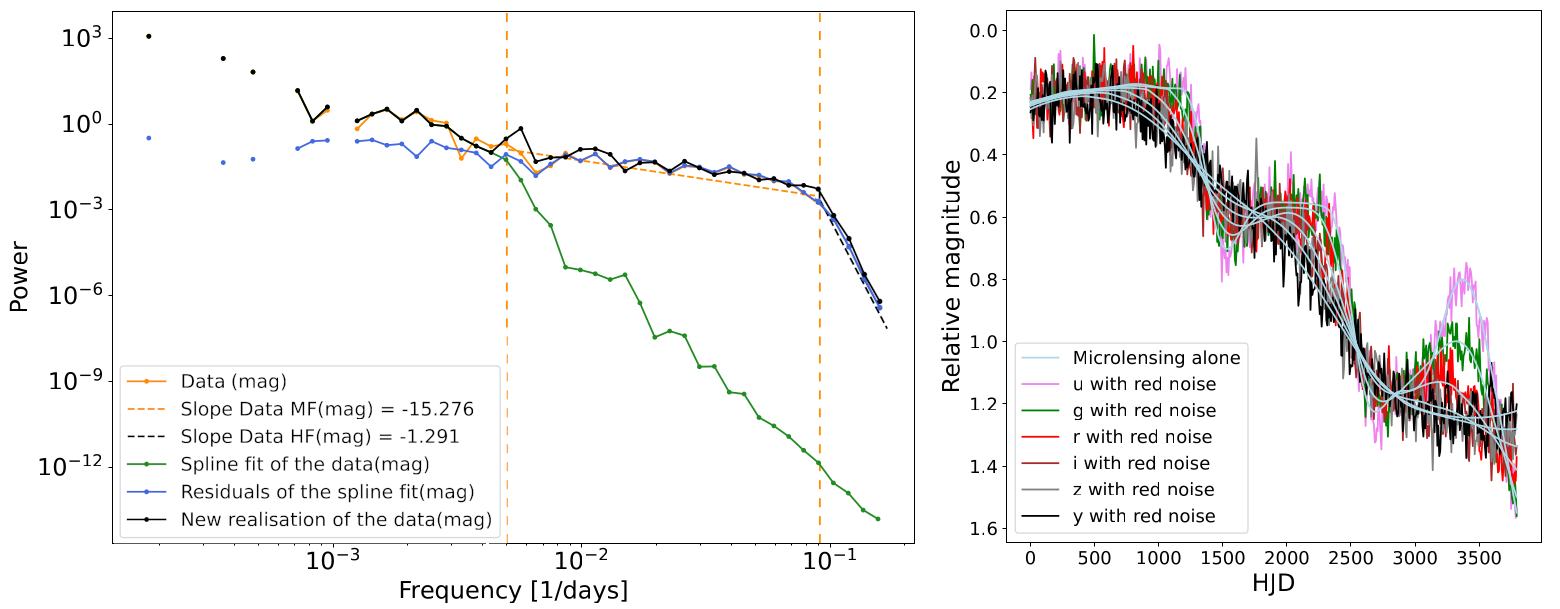}
    \caption{The left panel shows the power spectrum of the spline fit residuals shown in the bottom panel of Figure~\ref{fig:lc_data}. The vertical dashed lines at $1/200$ \invday and $1/11$ \invday delimit the low-, mid-, and high-frequency ranges. The slope of the red noise is separately fitted on the mid- and high-frequency ranges, as shown by the mid- (MF) and high-frequency (HF) slopes displayed. The red noise is then added on both the mid and high frequencies. The power spectrum of the residuals can be used to generate new realizations of the red noise, which can then be added to a simulated microlensing signal. The right panel shows an example microlensed light curve, with one realization of noise added with the same power spectrum as the left panel.}
    \label{fig:ps_fitting}
\end{figure*}

When generating the difference light curves, there may be uncertainty in measuring the time delay between images.
In addition, the reverberation of the continuum by the BLR with a time lag induces an echo of the intrinsic variability within the observed light curve. If the broad emission lines fall in the monitored band and the microlensing is not identical in both images, this echo will appear in the microlensing light curve~\citep[e.g.,][]{sluse14}. There may also be extended continuum emission coming from other regions of the disk~\citep{Sluse2024}.
The frequency and amplitude of this imprint are characterized by the size of the BLR and the variability of the continuum, which is well described by a damped random walk~\citep{macleod10}. 
Furthermore, the variability of the light curve (hence the microlensing light curve) can be affected by observational effects, such as the seasonal change of airmass and contamination by the lensed arc or the lensing galaxy when measuring the photometry of the images~\citep[see section 3.3.3 of][]{Sluse_2006}. 
Since these effects all require precise knowledge about the physical properties of the lens system, we choose to assimilate them into the red noise that is added to the immaculate microlensing light curve. 

For this work, we use data from the COSMOGRAIL program~\citep{courbin05}, which released the longest quasar microlensing light curves~\citep{millon20}, observed in the $r$ band with the Swiss 1.2 m Leonhard Euler Telescope. In particular, the microlensing light curve of \rxjeleven displays the events with the highest amplitude and the strongest correlated noise. Hence, we use these data to quantify a conservative red noise that can be added to LSST-like light curves.
As shown by Figure~\ref{fig:lc_data}, the microlensing light curve is first fit with a free-knot spline using the \pycs package \citep[see][for details of the implementation]{millonjoss}. These piecewise polynomials allow for a smooth fit of a targeted time scale of variation, by constraining the distance $\eta$ between two consecutive knots.   
Microlensing events are believed to occur on time scales longer than several years \citep[e.g.,][]{mosquera11}. We therefore choose $\eta = 300$ days to prevent the spline from fitting intraseason features while recovering the microlensing variations. The fit obtained is shown on the top panel in Figure~\ref{fig:lc_data}, and the residuals are shown in the lower panel. As shown in Figure~\ref{fig:ps_fitting}, the power spectrum of the data and the spline fit are identical for frequencies below $1/200$ \invday, which sets the boundary between high and low frequencies. At higher frequencies, the power spectrum of the data is identical to the residual one. Our goal is then to generate time series with the same power spectrum as the residuals to add to the simulated microlensing light curves. 
We use the red-noise generator implemented in \pycs to fit the slope $\beta$ and amplitude $\sigma$ of the residual power spectrum. To account for the change in slope of the power spectrum of the residual above $1/11$ \invday, we fit $\beta$ and $\sigma$ separately in the frequency windows [$1/200$,  $1/11$] \invday and $> 1/11$ \invday.  The addition of a generated red noise to the spline fit is shown in Figure~\ref{fig:ps_fitting} and has a power spectrum compatible with the original data. The same parameters are then used to add red noise to each simulated microlensing light curve.

\subsection{Training Labels}

We first identify all the peaks in the microlensing light curves of each lensed image that have a minimum amplitude of $0.5$ mag in the $u$ band, which is the bluest and thus most prone to microlensing variations.
This value is chosen to be roughly above the maximum amplitude of the added red noise.
After the difference light curves are generated and the red noise is included from Section~\ref{subsec:rednoise}, we apply an additional threshold to the identified HME peaks to avoid including HME peaks with little microlensing in the difference light curves, which could lead to our network predicting more false positives. We apply a threshold of 1 mag from the maximum to minimum of any time step in the light curve across all bands. If the light curve does not meet the threshold, then the microlensing events are not counted and the entire light curve is labeled with no HMEs. 
This threshold is applied in order to remove the initially identified peaks in the cases where the difference light curve shows little brightness variability.
We justify this because the typical length of the events is on the order of years, and the length of the light curves is set to 10 yr.
This makes it a common occurrence that two events from different light curves overlap, thus yielding a difference light curve that has less variability.
Once the HME event peaks are identified, the training label at each time step in the light curve is set to 1 if the time is within a 21 week window around each side of the peaks, or else it is set to 0. The 21 week window is chosen to be a bit more than half a season in length. The peaks in each difference light curve correspond to peaks in either image, so for a difference light curve A$-$B, peaks from image B will lower the relative flux and increase the relative magnitude.

\subsection{Baseline Observation Strategy} \label{sec:baseline_observing}

We first simulate microlensing light curves with a weekly cadence. We then want to degrade the cadences to mimic different observation strategies. Our baseline is LSST-like observations, which are simulated using the observation times produced by \texttt{rubin\_sim}\footnote{\href{https://github.com/lsst/rubin\_sim}{https://github.com/lsst/rubin\_sim}} and the \texttt{baseline\_v2.1\_10yr} cadences. We sample 100,000 different sky positions that have between 750 and 1000 total observations across the 10 yr to include only light curves
from the main LSST survey \citep[see section 2.4 of][]{Fagin_2024}. To ensure a consistent number of time steps across our light curves, we assign each observation to the nearest weekly interval. Times that are not observed in a band are masked by the RNN. Microlensing events happen over time scales of months to years, so combining the light curves to weekly intervals should not affect our ability to predict HMEs. Separate sky positions are used for the training, validation, and test sets to avoid bias.

\section{Machine learning model} \label{sec:ML}

\subsection{Neural Network Architecture} 

\begin{figure*}
    \centering
    \includegraphics[width=0.92\linewidth]{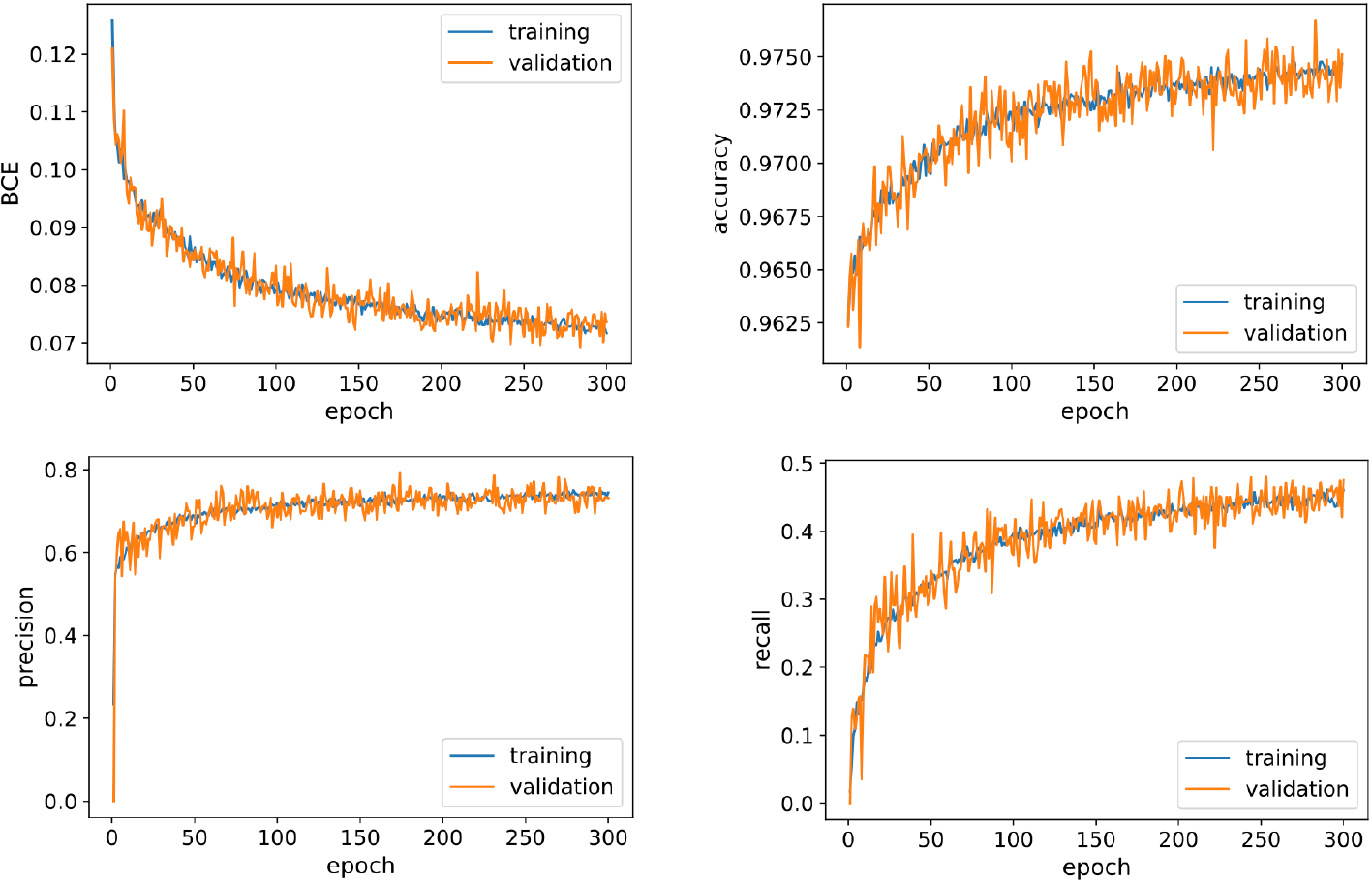}
    \caption{BCE loss (shown in the top left panel and given in Equation~(\ref{loss_function})), accuracy (top right panel), precision (bottom left panel), and recall (bottom right panel) as a function of the training epoch for the training (blue) and validation (orange) sets.}
    \label{fig:loss_vs_epoch}
\end{figure*}

Our neural network\footnote{\href{https://github.com/JFagin/Quasar_Event_Predictor}{https://github.com/JFagin/Quasar\_Event\_Predictor}} is trained to classify HME peaks of quasar microlensing light curves. The goal is to train our model so that it can be applied every week throughout the lifetime of LSST, including during seasonal gaps where there are no new observations. This is done by including only observations made at or before the current time of the prediction, then masking any further time steps. The network is trained to classify if there is an event peak within a 21-week time window before or after each time. We then make a prediction for each time step, starting when there are 85 weeks of data, so the network has at least two seasons or so of data to inform its predictions. We normalize the light curve by subtracting each magnitude by the first observation in the $r$ band (or the $g$-band for some observation strategies we test with no $r$-band observations). Data augmentation is used during training to avoid overfitting the training set. This involves selecting a random-cadence strategy for the light curve (i.e., for the nominal LSST light curves, we select the 10 yr cadence from a specific sky position) and a random time step of the observation. The validation and test sets remain fixed for comparison.

The input into the RNN is the relative brightness at each time step for each band. There are 522 time steps, corresponding to the brightness at each week across the ten yr. Nominally we include all six LSST bands, $ugrizy$, but we also explore situations where fewer bands are observed. Unobserved points are set to a dummy value of zero, to be masked by our network. We include an additional feature that is set to 1 up to and including the current time step, and then to 0 at later time steps. This feature is necessary to inform the network of the current time step, including when there are no new observations. 

Our RNN processes the time series in two separate paths, both forward and backward, since bidirectional RNNs can have improved performance. Each RNN path first contains a GRU-D layer~\citep{GRUD} a modified version of the gated recurrent unit~\citep[GRU;][]{GRU} that is designed to handle masking and irregular time intervals. This is followed by two GRU layers. Each GRU layer has a $\tanh$ activation function and a hidden size of 128. Each RNN layer is followed by layer normalization~\citep{layer_norm}. We also employ residual skip connections between the output of each RNN layer and the next layer, which have been shown to improve training stability and performance~\citep{ResNet}. The output of each RNN path is combined and followed by two fully connected (FC) layers. Each FC layer has a hidden size of 256 and is followed by a LeakyReLU activation function~\citep{maas2013rectifier} and layer normalization. The FC layers also include skip connections to the previous layer. We then have one final FC layer with a hidden size of 256 and an output size of 1, followed by a sigmoid activation function. The sigmoid activation function forces the final output of the network to be a value between 0 and 1, representing the probability of the HME peak being within 21 weeks of the time step. Each network has a total of 642,673 trainable parameters and is built using \texttt{PyTorch}~\citep{Pytorch}. 

\subsection{Neural Network Training}

The network is trained by minimizing a binary cross-entropy (BCE) loss function given by
\begin{equation}
\mathcal{L}(y,\hat{y}) = -\frac{1}{N}\sum_{i=1}^N y_i \log(\hat{y}) + (1-y_i) \log(1-\hat{y}_i) \, ,
\label{loss_function}
\end{equation}
where $y_i$ is the training label, $\hat{y}_i$ is the prediction of the network, and $N$ is the number of training examples. We minimize the loss function using an Adam optimizer~\citep{ADAM} with a batch size of 512 and a learning rate of 0.002 that is exponentially decreased to 0.0002 over the course of training. For training stability, we employ gradient clipping with a maximum gradient norm of 1. We train the network for multiple passes of our training set, known as epochs. We train for a total of 300 epochs. As mentioned previously, we use data augmentation to regenerate the training set each epoch by choosing a random-cadence strategy and time of the observation. We train separate networks for a variety of different observational strategies to see how the number of observations, seasonal gaps, and different bands affect the performance of our model. The BCE loss, accuracy, precision, and recall compared to the epoch are shown in Figure~\ref{fig:loss_vs_epoch} for the LSST baseline observations. We find further training beyond the 300 epochs to have very little improvement on the performance of our networks. The other networks, trained with the different observational strategies, show similar convergence after 300 epochs. Training each network took around four days with an NVIDIA A10 GPU (24 GB).

\subsection{Application to microlensing light curves} \label{sec:application_ML}

\begin{figure*}
    \centering
    \includegraphics[width=0.97\linewidth]{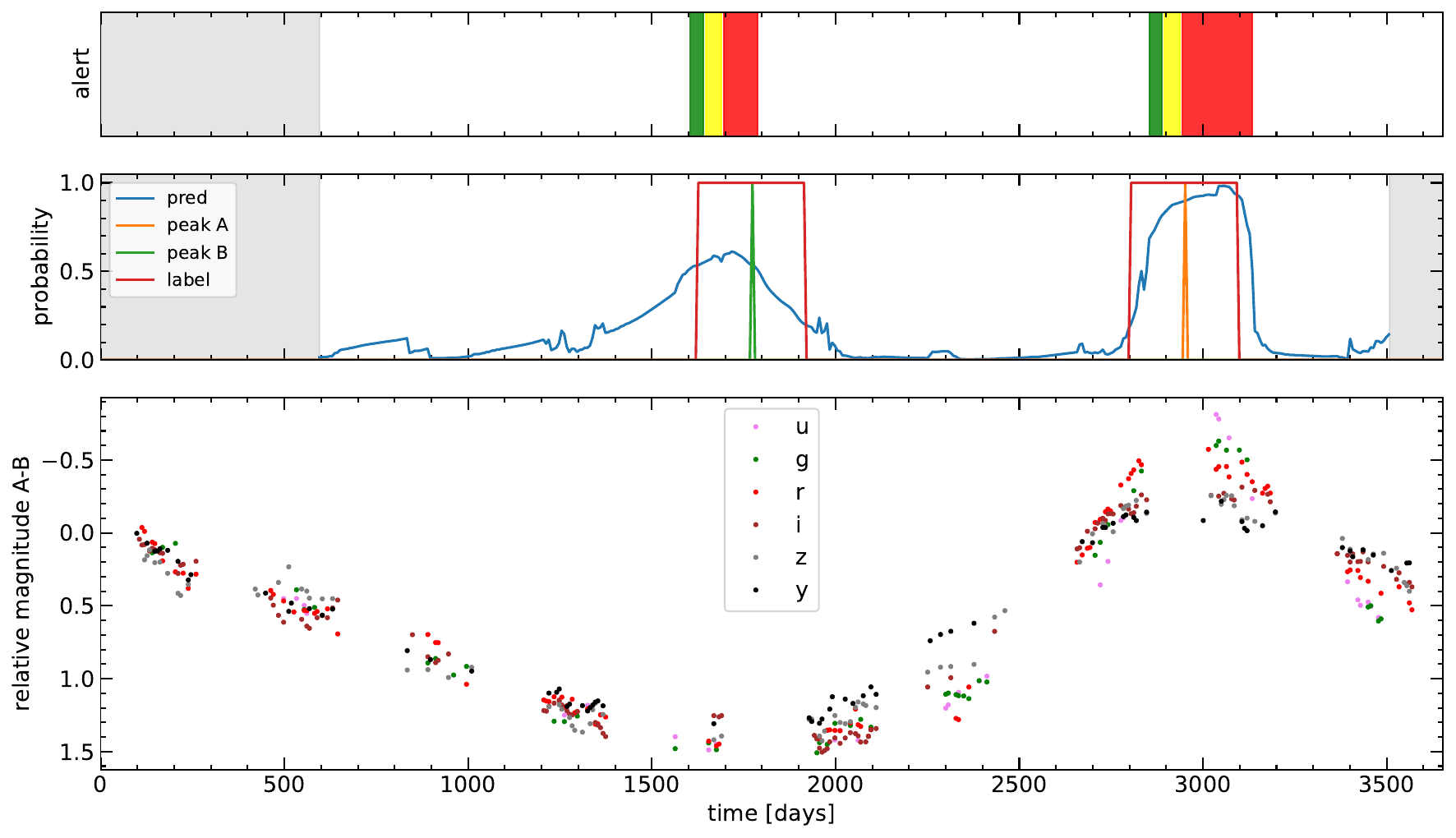}
    \caption{Example prediction of our network with LSST-like observations. The bottom panel represents the microlensing light-curve observations for the six LSST bands, normalized such that the first observation in the $r$ band is zero. The middle panel shows the output of our neural network by classifying if there is an event peak in either image within a 21 week window per side at each time step. The first 85 weeks and last 21 weeks are grayed out, since we do not make predictions for those times. The peak of the event is shown in orange and green for images A and B, respectfully, while the wider red label is the window that the network is trained to predict if there is an HME peak in either image. The top panel represents a mock alert system, where the first seven weeks of a continuous positive prediction (i.e., the probability is greater than 0.5 at the time step) are set to the green alert followed by seven weeks of the yellow alert and then greater than 14 weeks is the red alert.}
    \label{fig:prediction}
\end{figure*}

We propose a three-stage triggering system for rapid follow-up of HMEs. At each time step, the network makes a prediction to classify if an HME peak in either image is within a 21 week window of the latest data. We quantify a positive prediction if the network predicts a probability above 0.5. The three triggers are colored from lowest to highest alert as green, yellow, and red. The green alert is triggered after any positive prediction. The yellow alert is triggered if there is a green alert continuously for seven weeks in a row. The red alert is triggered if there is a yellow alert for seven weeks in a row and continues until a negative prediction. As soon as there is a negative prediction, the count is reset, to avoid false positives. Figure~\ref{fig:prediction} shows an example prediction of our network for an LSST-like light curve. In this case, our network is able to predict the HME peaks that occur in each image ahead of time. The network is still able to make predictions during the seasonal gaps, even though no additional observations take place. Ideally, the alert system should match up with the red label windows and we would predict the HME peaks 21 weeks beforehand, identifying their temporal locations. Another example is given in the Appendix that includes a false-positive, correct, and missed prediction of our network.

\section{Results} \label{sec:results} 

We explore a variety of observational strategies to assess how important high-cadence and multiband data are in predicting and analyzing HMEs. For each observational strategy, we train a separate network and then evaluate its performance on the test set. We use the same test set but with different observational cadences to fairly compare the performance of each model. Our baseline observing strategy represents LSST-like observations and is described in Section~\ref{sec:baseline_observing}. We also evaluate our model using a cadence strategy with no seasonal gaps in the data, where an observational mask is randomly sampled for each band at every time step. The mean number of visits across the 10 yr in the $ugrizy$ bands is [57, 72, 186, 194, 169, 174], taken from one of the baseline sky positions. We further consider a hybrid case in which we have LSST observations but a subset of bands is observed with the random-cadence mask to test how our network performs with partial observations during seasonal gaps. In addition, we evaluate the performance in the idealistic case, where we have regularly sampled data at every weekly time step. Moreover, we evaluate the performance when we have regular sampling but include seasonal gaps every year, which range from 140 to 180 days, chosen randomly for each light curve. For each cadence strategy, we additionally evaluate the performance using only a subset of the LSST bands. 

All the different strategies are summarized in Table~\ref{table:metrics} with various metrics to evaluate the performance of each model. The accuracy, precision, recall, and F1 score come from evaluating our network on random time steps throughout the survey across the test set. We note that the positive labels represent only about 3.6\% of the data set, so an accuracy of 0.964 could be achieved by just predicting the negative class. The correct peak, red in label, and false-positive metrics come from evaluating each time step of the entire 10 yr light curve for 1000 examples in the test set, i.e., making predictions as in Figure~\ref{fig:prediction} and weighted by the total number of HME peaks. The ``correct peak'' means the peak was correctly identified (i.e., the orange and green labels in the middle panel of Figure~\ref{fig:prediction} have blue predicted probabilities at least 0.5). ``Red in label'' means that there is a red alert within the 21 week event-peak window (i.e., the red label in the middle panel of Figure~\ref{fig:prediction} overlaps with the red alert in the top panel at some point). A ``false positive'' represents a red alert outside any event-peak window (i.e., if the red alert in the top panel of Figure~\ref{fig:prediction} did not overlap at all with the red label in the middle panel).

\begin{table*} 
\centering
\caption{Summary of metrics comparing the performance of each separately trained network on their test set using different observation strategies. The baseline for LSST-like observations is given at the top.}
 \begin{tabular}{c | c c c c | c c c} 
 Observation Type & Accuracy & Precision & Recall & F1 Score & Correct Peak & Red in Label & False Positive \\ [0.25ex] 
 \hline
 LSST, $ugrizy$ (baseline) & 0.976 & 0.764 & 0.466 & 0.579 & 0.551 & 0.571 & 0.204 \\
 LSST, $gi$ & 0.913 & 0.703 & 0.336 & 0.455 & 0.376 & 0.437 & 0.195 \\
 LSST $ugizy$, no seasonal gaps $r$ & 0.977 & 0.782 & 0.503 & 0.612 & 0.571 & 0.592 & 0.192 \\
 LSST $urzy$, no seasonal gaps $gi$ & 0.978 & 0.774 & 0.554 & 0.646 & 0.606 & 0.636 & 0.210 \\
 No seasonal gaps, $ugrizy$ & 0.979 & 0.799 & 0.596 & 0.683 & 0.688 & 0.717 & 0.216 \\
 Regular sampling with seasons, $ugrizy$ & 0.977 & 0.758 & 0.559 & 0.643 & 0.662 & 0.694 & 0.227 \\
 Regular sampling with seasons, $gi$ & 0.968 & 0.698 & 0.337 & 0.454 & 0.394 & 0.446 & 0.201 \\
 Regular sampling with seasons, $r$ & 0.967 & 0.720 & 0.137 & 0.230 & 0.157 & 0.192 & 0.122 \\
 Regular sampling, $ugrizy$ & 0.982 & 0.808 & 0.648 & 0.719 & 0.723 & 0.726 & 0.169 \\
 Regular sampling, $gi$ & 0.975 & 0.733 & 0.473 & 0.575 & 0.577 & 0.598 & 0.219 \\
 Regular sampling, $r$ & 0.968 & 0.630 & 0.200 & 0.303 & 0.239 & 0.233 & 0.143 \\
 \hline 
\end{tabular}
\label{table:metrics}
\end{table*}

Based on the different observational cadences we tested, we find the color information to be the most important factor for predicting HMEs. We expect this to be the case, since the different wavelengths probe different radii of the accretion disk and therefore are microlensed by different regions of the caustic map. With regular weekly observations in just the $r$ band, we are unable to make high-fidelity predictions, since we correctly identify just 24\% of the HME peaks. When we have regular sampling in both the $g$ and $i$ bands, the network performs much better, correctly identifying 58\% of the peaks. Additional bands further increase the performance of our model, but having at least two bands is most important.

We also find that the seasonal gaps play a significant role in the performance of our models. In the example in the Appendix, the missed HME peak at around 3200 days could have been predicted if the characteristic rise and band crossing did not fall within the season. Furthermore, the exact location of the HME peak at around 2000 days is not identified, because it falls directly in the seasonal gap. The network clearly knows there should be an HME peak somewhere in the seasonal gap but is unsure about its exact location. In addition, the irregular sampling can further hinder the ability of our network to accurately predict HMEs.
With all LSST bands but no seasonal gaps, we identified 69\% of the HME peaks instead of the LSST baseline case of 55\%. With regular sampling but including seasonal gaps, we identify 66\% of the peaks. With regular sampling and no seasonal gaps, this increases to 72\%. Thus, both the seasonal gaps and the irregular sampling can significantly hinder the performance of the model, although the seasonal gaps play a larger role. If we excluded identifying HME peaks that fall within the seasonal gaps, however, this would be less significant. We also show that some of the performance gap of the seasonal gaps can be made up if one or two bands are observed within the seasonal gaps.

\section{Discussion} \label{sec:discussion}

Our trained network is ready to be applied to the entire sample of lensed quasars of LSST. In the initial stages of the survey, before making predictions, we can fine-tune the network to calibrate it to the exact observational cadences and red-noise properties of the lensed quasar sample. It will also be necessary to measure the time delays between lensed images in order to produce the microlensing light curves. We begin making predictions after an initial time of 85 weeks, so the network has at least two seasons or so of data and to reduce the potential uncertainty in the time-delay measurements. A shorter initial time can be used if the time delays can be measured, although we would expect worse performance in the early time of the survey, since there are fewer observations to infer the microlensing time scales of the system. 

Our baseline LSST observations come from selecting sky positions within the main LSST survey. A limited number of strongly lensed quasars may be monitored in the Deep Drilling Fields, which will have much higher cadences. Our predictions would significantly improve from the increased number of observations. We expect the performance metrics for lensed quasars in the Deep Drilling Fields to be similar to the regular sampling with seasons in Table~\ref{table:metrics}.

Our goal is to train a general HME predictor for LSST. Thus, we use a diverse training set that should be representative of the entire expected LSST quasar sample. For specific systems that have already been well studied prior to LSST, we could train individual models to make more tailored predictions. For example, an HME predictor could be developed specifically for \qtwentytwo, like in~\citet{Wozniak_2000}. We would expect these models specially trained for individual systems to have improved performances, since they could specialize within a much narrower parameter space. We could also give the network priors on the $\kappa$, $\gamma$, $s$, and velocity distributions of systems if they have been estimated. Knowing these values could constrain the time scales of microlensing events and help the network identify where the peaks of the HMEs are with more clarity. The alert system itself could also be tailored to individual systems or populations with similar microlensing timescales. We choose a 21 week window around the HME peaks, so the training labels extend to around a season long. Some systems will be easier to trigger ahead of time, and we may then define a longer window, while others will be more difficult and may benefit from a narrower window. Fine-tuning the length of the training labels can be explored in future work.

In our microlensing simulation, we use a simple Gaussian brightness profile scaled to the disk size of each wave band. We expect that at the observational cadences of LSST, the shape of the inner region of the brightness profile will have little effect on the microlensing light curve. Furthermore, we are particularly interested in predicting the HMEs ahead of time. At these time scales and cadences, only the size of the disk chosen as its half-light radius is important, as the exact shape of the brightness profile has little impact on the light curve~\citep{Mortonson05}. This is no longer the case during caustic-crossing events, as the inner region of the accretion disk can be highly magnified. In such a case, a more realistic disk model should be used, including general-relativistic effects~\citep[e.g.,][]{best2024amoebaagnmodeloptical}. The analysis of the high-cadence follow-up, where the inner detail of the accretion disk becomes encoded in the light curve and may be able to constrain system parameters~\citep{best2022resolving}, is beyond the scope of this work. 

We build our training set by adding red noise from the residual power spectrum measured in the $r$ band of \rxjeleven, because this is currently one of the best-sampled, highest-signal-to-noise ratio light curves with well-measured time delays. Since the photometric Gaussian noise is fairly low, this system is ideal for isolating and quantifying the correlated red noise. Since the COSMOGRAIL data only have $r$-band observations, we assume the red noise in each wavelength to be equivalent. In reality, this might not necessarily be the case, depending on the physical origin of the red noise. If it originates from a flux contamination of the multiple image, we can expect a similar amplitude in all photometric bands. On the other hand, if it originates from the quasar variability echoed by the BLR, we may expect more correlated noise in the photometric bands overlapping with a broad emission line of the quasar. This would therefore depend on the redshift of the quasar. We choose a conservative approach for this work by using \rxjeleven, since it has one of the largest amplitudes of red noise. Once LSST data become available, the residual power spectrum of each LSST band can be measured from all monitored systems, and we can fine-tune our model based on the real population of red noise. The photometric errors are included in the red noise generated from \rxjeleven from the Leonhard Euler Telescope, which should be similar to LSST's photometric errors for objects this bright. Nevertheless, by using the entire LSST population, we will have access to more realistic photometric errors for each system. For bright quasars of ($<$$19$~mag), we expect the photometric errors to be only $\sim$$0.01$~mag and to play a minor role. For the fainter quasars ($>$$20$~mag), the photometric errors can be significant and our network should be adapted to better account for them. There may also be additional uncertainty captured within the red noise related to the uncertainty in measuring the time delay between lensed images. \rxjeleven is a well-studied system and there are only small uncertainties on its time-delay measurements. This will not be the case for all systems monitored in LSST, and this additional uncertainty will be incorporated into the population of red noise. Recent methods of image deconvolution could largely reduce these effects, hence reducing the amount of correlated noise in the extracted microlensing light curves~\citep{millon2024image}. In addition, the difference light curves need to be calculated when the observations in each lensed image overlap. This will cause the observation seasons to be shorter by a length of time equal to the time delays. For the majority of lensed quasars, the time delays will be of the order of weeks, and this should play only a minor role in our ability to predict HMEs. In data, we may focus follow-up efforts on lensed systems that have shorter time delays.

In our training set, we apply two thresholds to the HME labels. The first is that an event needs to be at least 0.5~mag in the $u$ band to be considered an HME.
This value is chosen to be just above the maximum amplitude of the added red noise.
Furthermore, in the subtracted microlensing light curves, we apply an additional threshold of 1~mag between the maximum and minimum of the light curve across all bands. This second threshold helps exclude labels where long-term microlensing events cancel each other out, leaving little indication of an HME in the light curve.
These thresholds are designed to minimize true false positives and avoid triggering expensive follow-up observations unnecessarily. Moreover, these thresholds encourage follow-up to be triggered for only the strongest HMEs, when microlensing \textit{zooms in} on the inner regions of the disk. Indeed, the choices of these thresholds will affect how the network performs. In the Appendix, we show an example where if the threshold had been lower, the first false positive would have been considered a correct prediction, as this peak would have been labeled. Thus, our reported false-positive rate is a conservative estimate, since edge cases where the HME is just below the threshold will often be misclassified by the network. Our focus is on predicting the HMEs that have the most significant impacts on the microlensing light curves. Future studies could explore adjusting these thresholds, potentially lowering them to capture more subtle events.
This approach could be particularly useful in scenarios where the amplitude of the red noise is reduced, allowing for a broader range of HMEs to be detected. We could also train multiple networks with different thresholds in the training set and use the ensemble of different predictions to get a measurement of the strength of the HME.

Ideally, we would simulate the intrinsic quasar variability and use a variable disk when convolving with the microlensing maps, instead of using a data-driven approach. Then we would measure the time delays between each image using the initial simulated data of LSST and subtract the images to produce our simulated microlensing light curves. This would then include the effects from the uncertainty in measuring the time delays, the photometric errors, and the red noise in a more self-consistent way. We would, however, need to model the BLR reverberation, which can have very complicated geometry. There may also be extended continuum emission coming from other regions of the disk that we may not be able to simulate~\citep{Sluse2024}. Since accounting for all the aspects of the intrinsic variability may be infeasible, we prefer a data-driven approach and defer exploring the inclusion of these effects directly into the simulation for future work.

Here, we fix the threshold for a positive prediction of our network to a probability of 0.5, as is typically done in binary classification. By raising this threshold, however, we would predict fewer HME peaks but lower the false-positive rate. We could also adjust the balance of classes in the training set. During training, we mimic observing at a random time throughout the 10 yr survey, and we find a rate of positive labels to be around 3.6\%. We could instead choose to bias the training set, by including more or less mock observations near the HME peaks. Including more time steps near the HME peaks would enable the network to make more positive predictions, increasing the number of correctly predicted HME peaks but also raising the false-positive rate. Within our metrics, we include predictions that fall within seasonal gaps. By focusing low-cost photometric follow-up near the predicted HME, we could improve our prediction and reduce the number of false positives before triggering more expensive follow-up observations, such as X-ray or spectroscopic monitoring.

We may also improve upon our machine learning model in future work. We use RNNs that include the masking of missing values through GRU-D layers~\citep{GRUD}. We may also explore other RNN architectures, such as stochastic RNNs, which have been used to model UV/optical quasar variability~\citep{Sheng_2022}. Continuous recurrent units~\citep{schirmer2022modeling}, which are a probabilistic recurrent architecture where the hidden state evolves as a linear stochastic differential equation, are a promising way of dealing with the irregular sampling. We may also consider using latent stochastic differential equations, which have worked well at modeling the intrinsic quasar variability for simulated LSST light curves~\citep{Fagin_2024,fagin2024jointmodelingquasarvariability}. Transformers~\citep{vaswani2023attention} such as multi-time attention networks~\citep{shukla2021multitime}, where the time embedding is fed to the attention mechanism to incorporate irregularly sampled time series, may also show improved performance compared to RNNs.

The largest current catalog of monitored lensed quasars is COSMOGRAIL, which has only $r$-band observations~\citep{millon20}. We show in Table~\ref{table:metrics} that for regular observations including seasonal gaps in the $r$ band, only around 15.7\% of event peaks are correctly identified, with a false-positive rate of 12.2\%. These metrics indicate that even with the most ideal case of regular weekly sampling, without multiband data the network fails to make meaningful predictions. There is currently no catalog of long-term monitored lensed quasars with multiband data, so we defer testing our model on real data for future work.

\section{Conclusions} \label{sec:conclusion}

We trained an RNN to predict HME peaks in LSST-like microlensed quasar light curves. We make our model and training procedure open-source and available at \href{https://github.com/JFagin/Quasar_Event_Predictor}{https://github.com/JFagin/Quasar\_Event\_Predictor}. A pretrained neural network like ours could be quickly and autonomously applied to the entire sample of monitored LSST strongly lensed quasars each week as new data are obtained, in less than a minute on a single GPU. This is the first deep-learning method of predicting HMEs. 

For the baseline LSST cadences, we can expect to correctly identify around $\sim$55\% of HMEs, with a false positive rate of $\sim$20\%. We therefore expect to correctly identify tens to hundreds of HMEs per year~\citep{Neira20,Taak_2023}. These metrics can be improved as high-cadence follow-up is triggered ahead of the peak of the event. 

We envision our network being continuously applied throughout the entire LSST survey to give weekly notifications to experts on potential incoming HME peaks. These alerts will be crucial for optimizing follow-up observational resources. High-cadence multiband or spectroscopic follow-up observations of these HMEs will offer unique insight into the inner structure of AGNs as imprinted in the light curves during caustic-crossing events. 

\software{
\texttt{PyTorch}~\citep{Pytorch},
\texttt{Numpy}~\citep{Numpy}, 
\texttt{Matplotlib}~\citep{Matplotlib},
\texttt{Astropy}~\citep{astropy:2013,astropy:2018,astropy:2022}
}

\section*{Acknowledgments}

This work originated in the Lensing Odyssey 2021 workshop, and so we would like to acknowledge the organizers and attendees for the fruitful discussions. We thank the anonymous referee for the useful comments. We also thank Padma Venkatraman for useful discussions. This work used resources available through the National Research Platform (NRP) at the University of California, San Diego. NRP has been developed, and is supported in part, by funding from the National Science Foundation, from awards 1730158, 1540112, 1541349, 1826967, 2112167, 2100237, and 2120019, as well as additional funding from community partners. Support was provided by Schmidt Sciences, LLC. for J.F., H.B., M.O., and G.V. M.M. acknowledges support by the SNSF (Swiss National Science Foundation) through mobility grant P500PT\_203114 and return CH grant P5R5PT\_225598. T.A. acknowledges support from ANID-FONDECYT Regular Project 1240105, ANID Millennium Science Initiative AIM23-0001 and the ANID BASAL project FB210003. H.B. acknowledges the GAČR Junior Star grant No. GM24-10599M for support.

\appendix

\section{Another example prediction} \label{sec:appendix}

\begin{figure*}
    \centering
    \includegraphics[width=0.85\linewidth]{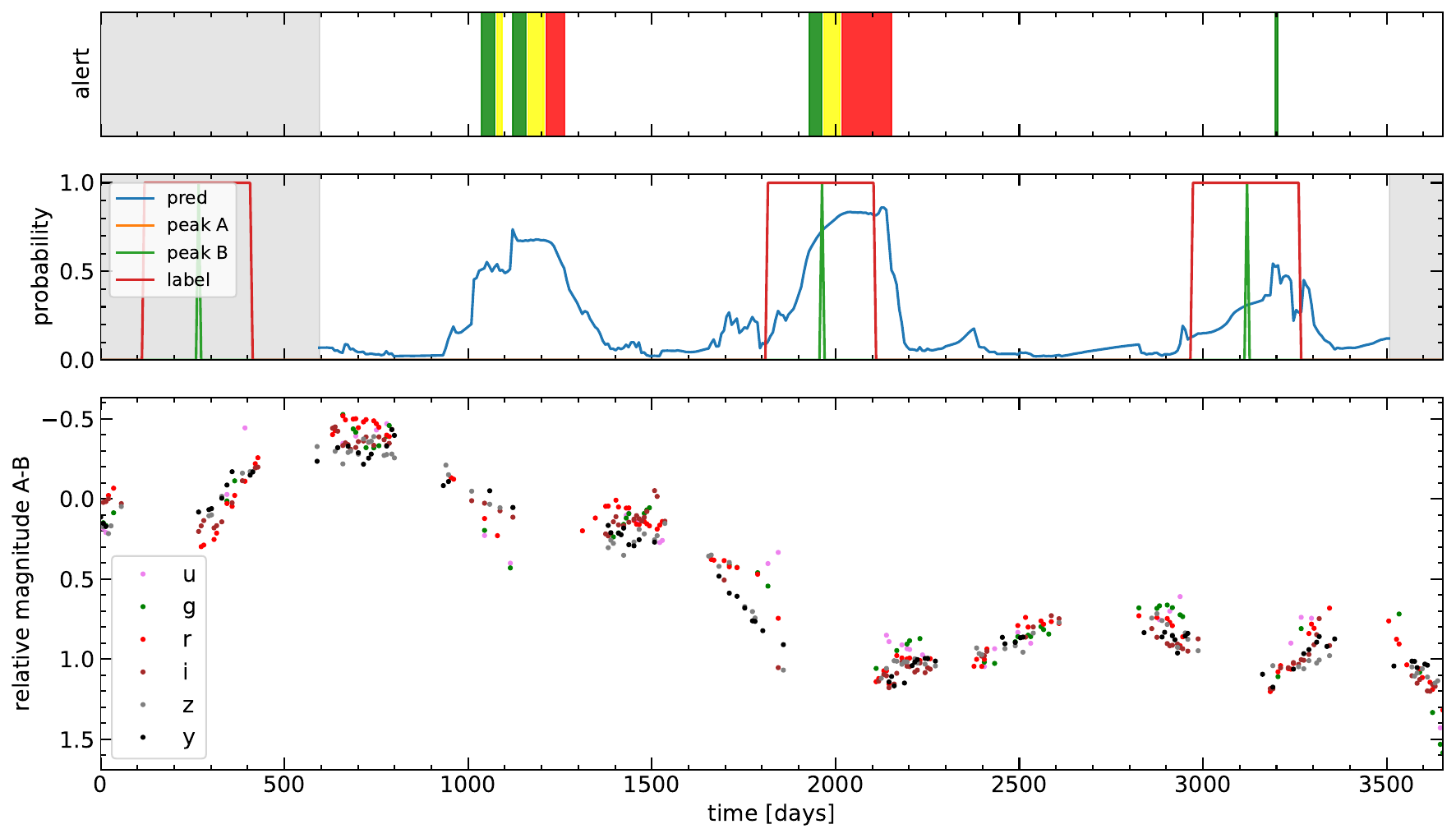}
    \caption{Example of the prediction of our network on LSST-like observations like Figure~\ref{fig:prediction}.}
    \label{fig:prediction_appendix}
\end{figure*}

In Figure~\ref{fig:prediction_appendix} we show another example of the predictions of our network. In this case, there is a false positive, correct prediction, and missed prediction, from left to right.

\bibliography{bib}{}
\bibliographystyle{aasjournal}

\end{document}